\documentstyle[prl, aps, epsf, preprint]{revtex}
\begin{document}
\draft
\preprint{Y123/Pr123-v1.4}

\title{
$c$-Axis tunneling in YBa$_2$Cu$_3$O$_{7-\delta}$/PrBa$_2$Cu$_3$O$_{7-\delta}$
superlattices
	}
	
\author{
	J.C.Mart\'\i nez, A.Schattke, M. Jourdan, G.Jakob, H.Adrian
	}
	
\address{
	Johannes Gutenberg -University of Mainz; Institute of Physics; 55099
Mainz; Germany
	}
	
\date{ Feb. 17, 1999}

\maketitle

\begin{abstract}
In this work we report $c$-axis conductance measurements done on a superlattice based on 
a stack 
of 2 layers YBa$_2$Cu$_3$O$_{7-\delta}$ and 7 layers PrBa$_2$Cu$_3$O$_{7-\delta}$ 
(2:7). We find that these quasi-2D structures show no clear superconducting coupling 
along the 
$c$-axis. Instead, we observe tunneling with a gap of $\Delta_c=5.0\pm 0.5$ \ meV for the 
direction perpendicular to the superconducting planes. The conductance spectrum show well 
defined quasi-periodic structures which are attributed to the superlattice structure. 
From this 
data we deduce a low temperature $c$-axis coherence length of $\xi_c=0.24\pm 0.03$\ nm.
\end{abstract}

\pacs{74.80.Dm,73.61.-r,73.20.Dx}


%

As for classical superconductors, tunneling experiments are a direct way of testing the 
local 
superconducting density of states \cite{giaevaer}. In particular the co-existence of $s$-
wave and 
$d$-wave components of the order parameter was directly investigated from $c$-axis planar 
tunneling measurements done in high temperature superconductors (HTS) \cite{sun,kleiner}. 
However tunneling experiments are extremely sensitive to the quality of barriers and 
interfaces. 
This aspect explains the extreme care taken by the different groups in the fabrication of 
HTS 
tunnel junctions. One way of getting around this problem is to investigate $c$-axis 
tunneling in 
YBa$_2$Cu$_3$O$_{7-\delta}$/PrBa$_2$Cu$_3$O$_{7-\delta}$ superlattices (Y123/Pr123). 
In those systems it is possible to modify the tunneling properties simply by varying the 
periodicities of the Y123 and Pr123 layers. Another advantage is that transmission 
electron 
microscope (TEM) studies show atomically flat Y123/Pr123 interfaces in superlattices 
\cite{jia}.

In this work we study the influence of the periodicity of an artificial superlattice on 
the local 
superconducting density of states of Y123. This was only possible by a sustained effort 
divided 
in three main steps: the preparation of high quality Y123/Pr123 superlattices, the 
patterning of 
suitable mesa structures and the measurement of the $c$-axis transport properties.

%

Y123/Pr123 superlattices are deposited on a similar way as Y123 thin films. The only 
difference is 
that after sputtering one to 10 unit cells (u.c.) of Y123 the substrate is turned to a 
cathode 
containing the Pr123 target. This process is repeated until a total film thickness of 
200\ nm. The 
switching between targets is made with a computer-controlled step motor. To provide low 
ohmic contacts the process ends by the in-situ deposition of a protective gold layer with 
thickness from 200 to 400 nm. In order to avoid the formation of pin-holes and reduce the 
surface roughness the sputtering process was done at 3\ mbar pressure of a mixture 1 to 2 
of 
O$_2$ and Ar. 

The crystallographic quality of the superlattices used in this work has been checked by 
x-ray 
diffraction. Our measurements showed up to third order satellite peaks observed in 
($\theta -
2\theta$) scans. On the other hand TEM studies show steps of only 1\ u.c. for every 100\ 
nm. 
This value is much smaller than the 7 u.c. thick Pr123 barrier. 

Given that a Gold layer covered the superlattices the surface morphology could not be 
checked 
directly on the same samples where the $c$-axis measurements were done. However scanning 
tunneling microscopy done in similar superlattices reveal an average surface modulation 
of 
$\pm$7.2 nm (6 u.c.) for a length scale of 1.6 $\mu$m. This roughness is very small when 
compared to the total thickness of the mesas which was of about 120 nm. Since the 
measurements below were performed on 2:7 superlattices and the top layer was always Y123, 
only the two upper Y123 layers (of a total of 10) are probably affected by the protective 
gold 
layer. This is expected to have little influence on our experimental results.

Before making the mesa structures, a ground electrode with 1.2 $\times$ 10 mm$^2$ was wet 
chemically etched. The mesas were later on prepared by standard UV-photolithography and 
ion 
milling. During the etching process the samples were cooled down to 77\ K. Later the chip 
was 
coated with photoresist, and a window was opened on the top of each mesa by a 
photolithographic process. The preparation step ends with the deposition of a gold top 
electrode patterned by wet chemical etching. Because of its smoothness, high homogeneity 
and 
low defect density the photoresist was directly used for insulating the top contact from 
the 
ground electrode. All measurements were done in a ''three point'' geometry, where the 
typical 
contact resistance between the HTS material and gold is $R_S \approx 3\times 10^{-5} 
\Omega$cm$^2$. For the moment, we are limited to mesa structures down to 15$\times$15 
$\mu$m$^2$.

%

\begin{figure}[t]
	\centering
	\epsfxsize = 12 cm
	\epsfbox{
		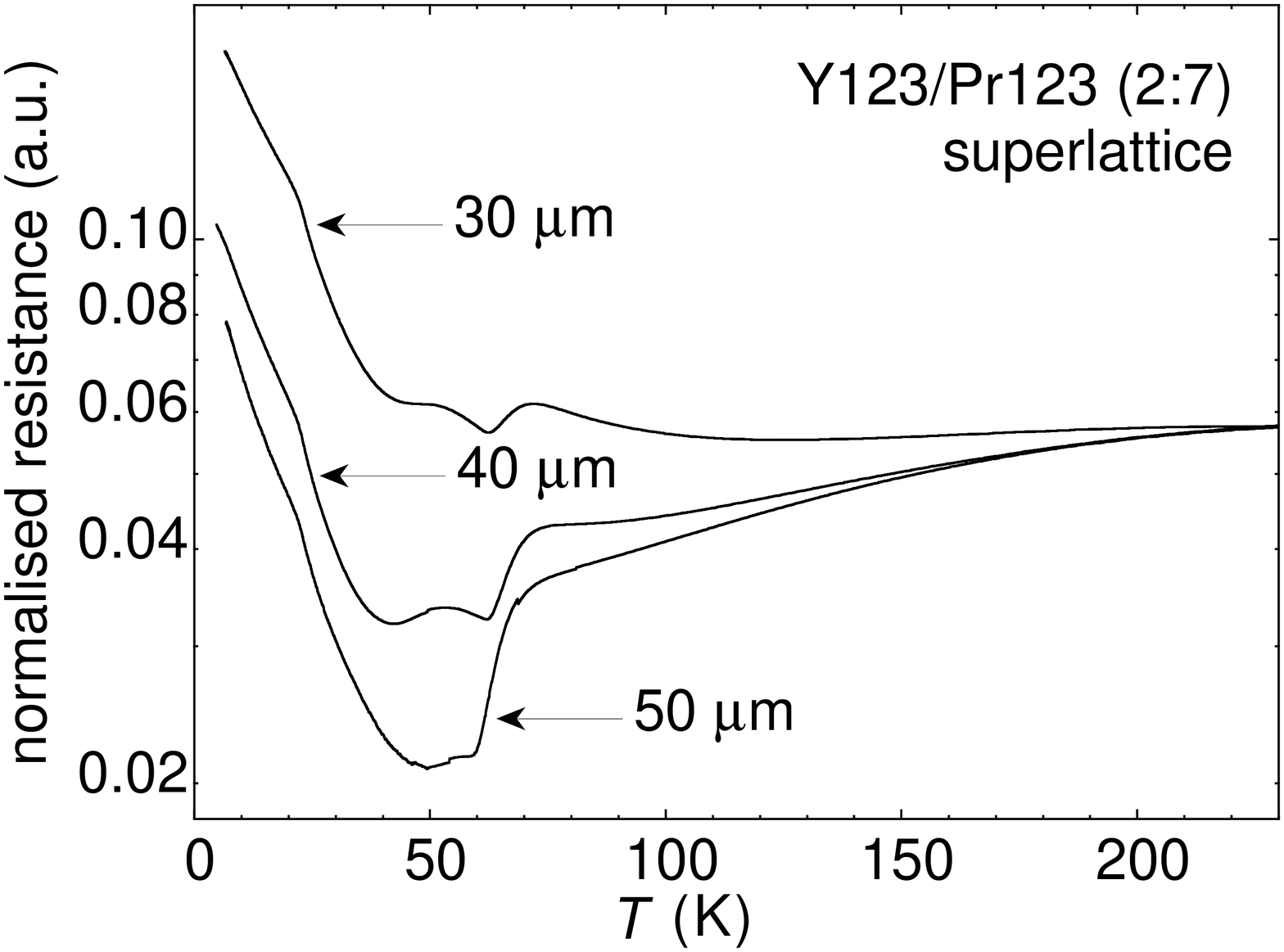}
	\vspace{5 mm}
	\caption{
Resistance of $30\times 30$, $40\times 40$ and $50\times 50$\ $\mu$m$^2$ mesas patterned 
on a 
2:7 Y123/Pr123 superlattice. The three measurements are normalized to the resistance at 
250\ K 
and represented with a y-logarithmic axis. \label{fig1}}
\end{figure}

In Fig.\ \ref{fig1} we show a semi-logarithmic plot of the resistances of three mesas 
with $30\times 
30$, $40\times 40$ and $50 \times 50\ \mu$m$^2$ prepared on a Y123/Pr123 superlattice 
with 
2 layers Y123 and 7 layers Pr123 (2:7). The transition observed at 65\ K corresponds to 
the 
superconducting transition $T_c$ of the Y123 layers. The reduced $T_c$ is typical for the 
non-fully developed order parameter in the 2 u.c. thick Y123 layer \cite{jakob}. Above 
$T_c$, 
the larger mesas show a temperature dependence that is similar to the one measured in the 
(a,b)-plane. This demonstrates that we measured a non-negligible amount of the (a,b) 
component. However below $T_c$ the superconducting Y123 layers define equipotential 
planes, and only the $c$-axis component of the resistance can be measured. This is 
confirmed by 
the vertical shift existing between the different plots shown in Fig.\ \ref{fig1}. The 
logarithmic y-
axis shows that the different curves differ below $T_c$ only by a proportionality factor. 
The 
resistances of the mesas at 20 K were $R_c(30)=168$\ $\Omega$, $R_c(40)=68$\ $\Omega$ and 
$R_c(50)=63$\ $\Omega$ which give ratios of $R_c^{50}/R_c^{30}=0.38$ (expected 
$50^2/30^2=0.36$) and $R_c^{40}/R_c^{30}=0.40$ ($40^2/30^2=0.56$). The discrepancy of 
30\% observed in the $R_c^{40}/R_c^{30}$ can be attributed to a larger degree of damage 
introduced during the etching of the $40\times 40$\ $\mu$m mesa. The contact resistance 
of the 
gold top electrode should scale as well with the area of the mesas.

%
%

\begin{figure}[t]
	\centering
	\epsfxsize = 12 cm
	\epsfbox{
		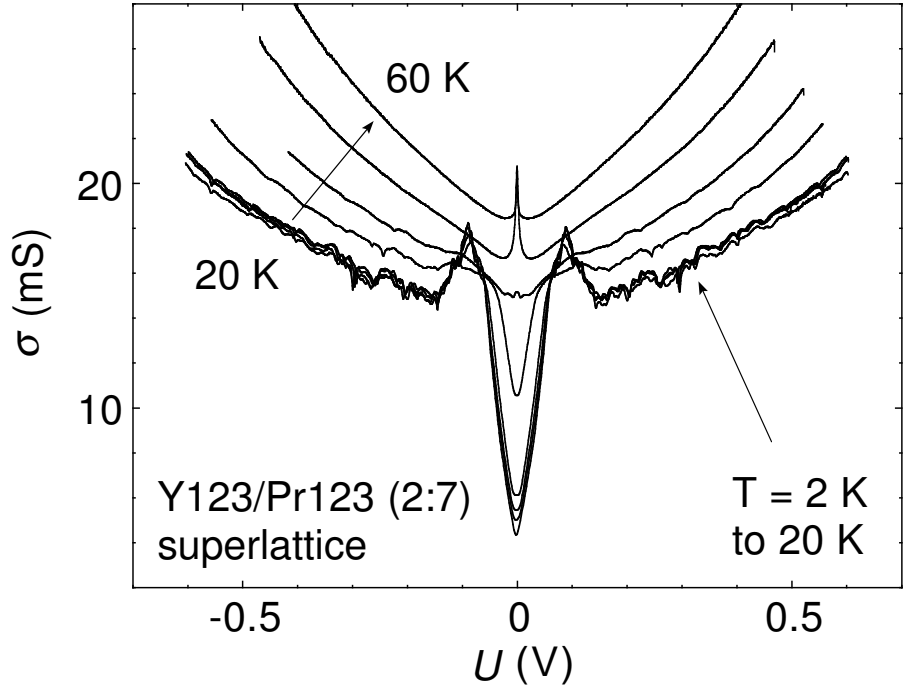}
	\vspace{5 mm}
	\caption{
Differential conductivity $\sigma(U)$ of a $30\times 30 \mu$m$^2$ mesa for temperatures 
between 
2.0 and 60\ K. The parabolic background can be associated to a barrier with $3.5$ \ nm 
width and 350 meV height (see text).
\label{fig2}}
\end{figure}

We measured simultaneously the $U$ vs. $I$ characteristics and the differential 
resistance in the 
$30\times 30$\ $\mu$m$^2$ mesa. This was done by using a battery operated current source 
that superposes above an arbitrary DC current, a small AC signal generated by the 
reference of a 
Lock-in amplifier. The DC-and AC-signals were measured with an HP34420A nano-voltmeter 
and a PAR 5210 Lock-in amplifier respectively. In Fig.\ \ref{fig2} we show some of our 
results 
on a $30\times 30 \mu$m$^2$ mesa done on a 2:7 superlattice for temperatures between 2.0 
and 
60\ K. Several features can be identified in Fig.\ \ref{fig2}. The first is the parabolic 
background 
which can be well described by the Simmons model which predicts \cite{simmons}: 
\begin{equation}
\sigma_b(U)=\sigma_0\left( 1+\zeta U^2\right)
\label{eq1}
\end{equation}
This model corresponds to a metal-insulator-metal junction with a rectangular barrier of 
width $d$ 
and height $\Phi$. The constant factors are $\sigma_0=(3/2)(e(2m\Phi)^{1/2}/h^2d) \exp(-A 
d 
\Phi^{1/2})$ and $\zeta = (Aed)^2/96\Phi$ where $A=4\pi(2m)^{1/2}/h$. Considering that we 
have 10 bi-layers connected in series, we deduced from Eq.\ \ref{eq1} that each Pr123 
barrier has 
below $T=25$\ K an effective height $\Phi = 370$\ meV and an effective width $d=3.5$\ nm. 
Although $d$ is much smaller than the 8.2\ nm expected from the $7$\ u.c., we can explain 
this 
result by pure geometrical arguments. If we take into account the imperfections in the 
superlattice (steps of 1 u.c. per interface) and the fact that superconductivity extends 
to the 
chains (0.5 u.c. per interface), we would have an effective barrier thickness of $\approx 
4.8$\ nm 
(4 u.c.). This crude estimation is only $37\%$ larger than $d$.
From our data we deduce that the values of $d$ and $\Phi$ were practically temperature 
independent up to $25$\ K. Above this temperature the conductivity follows the behavior 
characteristic for resonant tunneling via up to two localized states \cite{glazman}:
\begin{equation}
\sigma_b(U)=g_0+\alpha U^{4/3}
\label{eq3}
\end{equation}

The two peaks observed at lower temperatures in $\sigma(U)$ should indeed correspond to a 
$c$-axis superconducting gap. The distance between the two peaks is $U_{pp}=178$\ mV. 
This 
particular mesa was made with a height of about 120 nm estimated from an etching rate 
calibrated by Atomic Force Microscopy done in different etched films. This gives a stack 
of 
$n=8$ to $10$ bi-layers which present each a $c$-axis superconducting gap of 
$\Delta_c=U_{pp}/4 n=5.0\pm 0.5$ \ meV.

%
%

This value is in excellent agreement with the value of $\Delta_c$ given in the literature 
which 
scatters between 4 and 6 meV for planar junctions prepared with Pr123, CeO$_2$ and 
SrAlTaO$_6$ barriers \cite{iguchi,ying,bari,nakajima}. To understand this result we have 
to 
look at Scanning Tunneling Spectroscopy (STS) data. Typical STS measurements done on both 
Y123 thin films and high quality single crystals give gap structures at about 5 and 20 
meV 
\cite{miller,maggio}.These fine structures observed in tunneling spectra were explained 
successfully by Miller at al by considering that Y123 is constituted by a stack of strong 
and 
weak superconducting layers which contribute with different weights to the tunneling 
spectra 
\cite{miller,tachiki2}. In particular the 5 meV gap has been attributed to the BaO and 
CuO layers 
situated between the CuO$_2$ blocks. Since these BaO and CuO layers are common to both 
Y123 and Pr123 we expect them to constitute the interfaces which are relevant to our 
tunneling 
process. Given that $T_c$ in our superlattices is only reduced by 20\% we do not expect 
the 
CuO$_2$ superconducting gap to be strongly suppressed. On the other hand given that a 4-6 
meV gap structures are observed in measurements done with different barriers we think 
that the 
gap $\Delta_c\approx 5$\ meV is indeed an intrinsic property of Y123. The existence of 
this small c-axis gap is consistent with the strong thermal smearing of the gap feature 
at about 50\ K, which corresponds to an energy of the same magnitude as $\Delta_c$. 

For the moment it is not clear to us 
which will be the influence of the CuO chains to the symmetry of the order parameter. 
However 
since these chains form together with the apical oxygen CuO$_2$ cells oriented along to 
the $c$-
axis, it is likely that a $c$-axis gap would exist in Y123 with a $s$-wave like symmetry.

%
%

\begin{figure}[t]
	\centering
	\epsfxsize = 12 cm
	\epsfbox{
		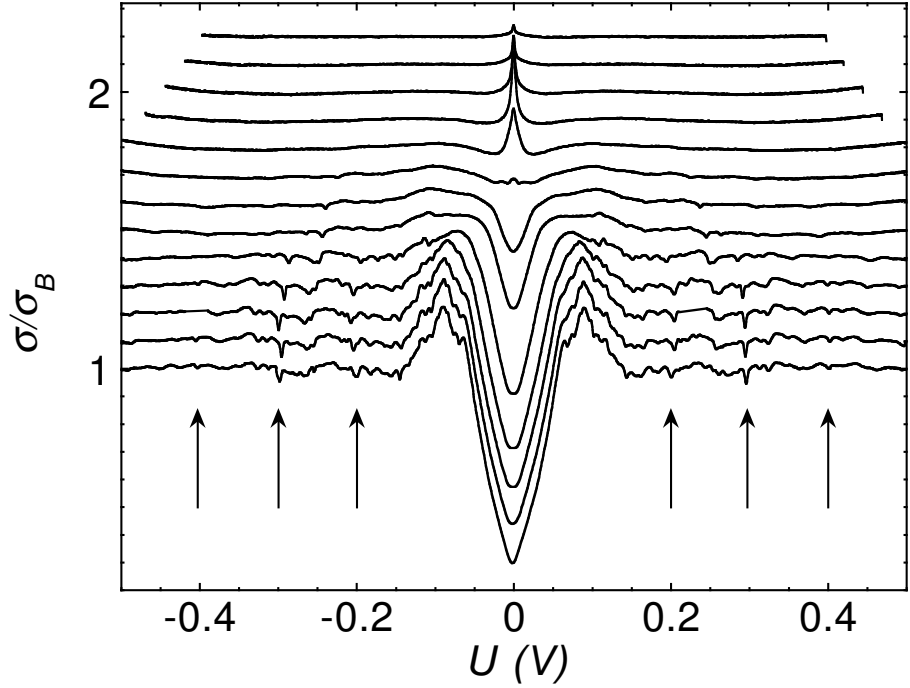}
	\vspace{5 mm}
	\caption{
Differential conductivity $\sigma$ divided by the parabolic background $\sigma_b$ for 
2.0, 10, 15, 
20, 25, 30, 35, 40, 45, 50, 55, 60 and 65\ K (from bottom to top). The different 
measurements 
are shifted vertically for sake of clarity. The arrows indicate sharper features 
noticeable in the 
tunneling spectrum.
\label{fig3}}
\end{figure}

To enhance the other features present in $\sigma(U)$, we plot in Fig.\ \ref{fig3} 
$\sigma/\sigma_b$ 
for temperatures between 2.0 and 65\ K. For more clarity the data is vertically shifted. 
Below 
25\ K the Simmons model is used to calculate $\sigma_b$ (Eq.\ \ref{eq1}). Above that 
temperature the resonant tunneling expression given by Eq.\ \ref{eq3} is employed. We 
would 
like to emphasize that for all temperatures the low bias features were included in the 
estimation 
of $\sigma_b(U)$. 

Below 30\ K we observe a number of reproducible features which were superposed to the 
superconducting gap. For increasing temperatures, the gap and these structures are 
smeared out 
by thermal fluctuations. At about 30\ K the only signature of the gap is a soft voltage 
dependence of $\sigma$ and a zero bias peak which starts to develop, grows up to about 
50\ K 
and finally disappears near $T_c$.

%
%
%
%

Although there is not for the moment a clear explanation for the origin of this zero bias 
peak, its 
disappearance close to $T_c$ shows that it is clearly related to tunneling of 
superconducting 
pairs. As suggested by Abrikosov \cite{abrikosov} this anomaly could be due to resonant 
tunneling through localized states in the barrier. The absence of a zero 
bias peak at lower temperatures is consistent with this picture: the success in fitting 
the data 
with a Simmons model indicates that below 25\ K, the Pr123 layers behave 
predominantly like normal tunneling barriers having one or no resonant states.

\begin{figure}[t]
	\centering
	\epsfxsize = 12 cm
	\epsfbox{
		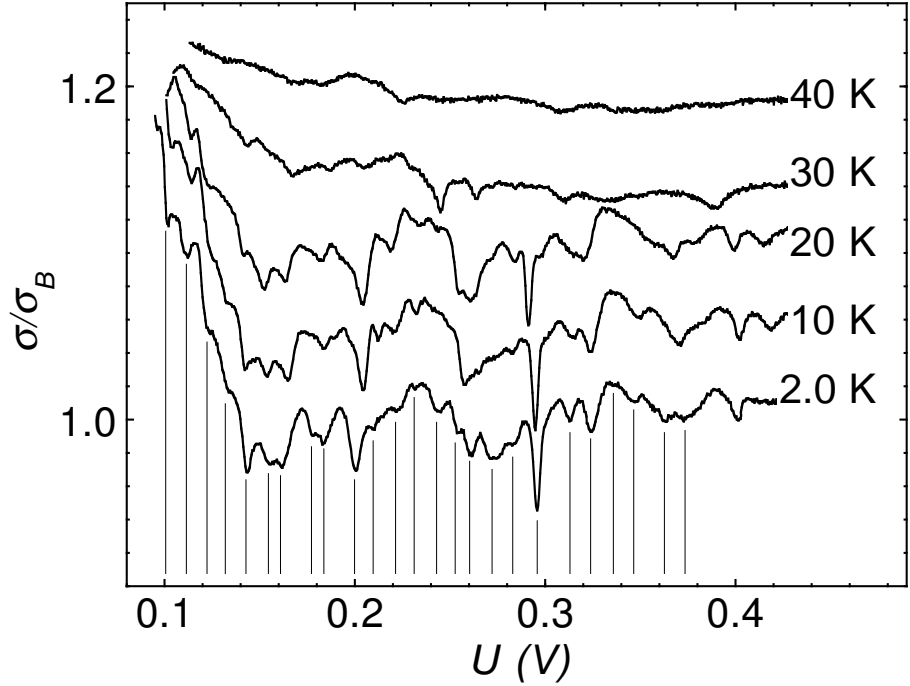}
	\vspace{5 mm}
	\caption{
Same data as in Fig.\ 3 plotted on a larger scale for the indicated temperatures. The 
vertical lines 
indicate the minima between 0.1 and 0.3\ V observed in $\sigma/\sigma_b$ for $T=2$\ K.
\label{fig4}}
\end{figure}

\begin{figure}[t]
	\centering
	\epsfxsize = 12 cm
	\epsfbox{
		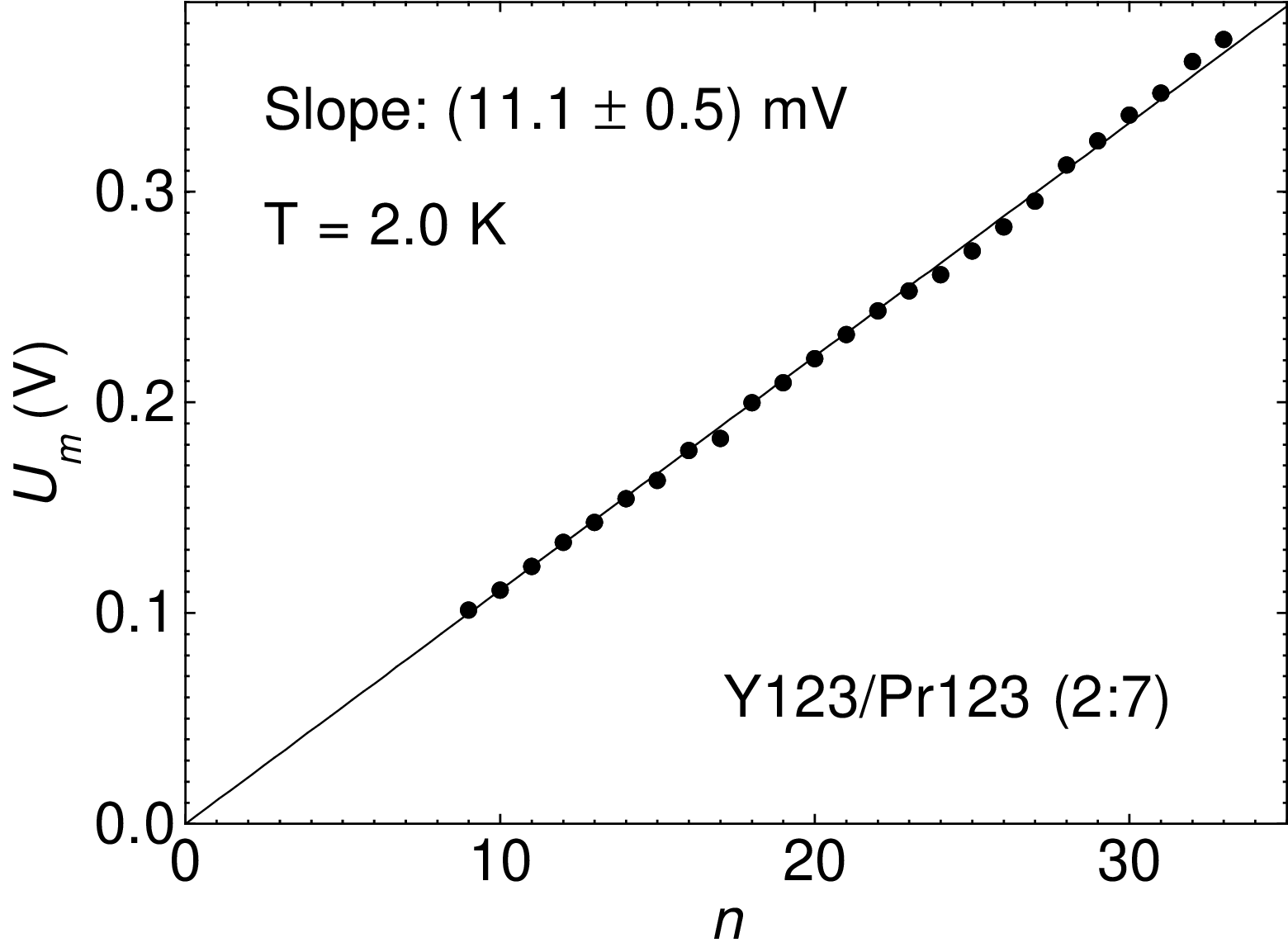}
	\vspace{5 mm}
	\caption{
Minima extracted from $\sigma/\sigma_b$ plotted as function of an integer index $n$. The 
slope of 
$(11.1 \pm 0.5)$\ mV corresponds to a quasi-periodicity of the smaller structures in 
Fig.\ 
\ref{fig4}.
\label{fig5}}
\end{figure}

The arrows in Fig.\ \ref{fig3} show sharper indentations in $\sigma(U)$ which can be 
followed up 
to 30\ K. To investigate the additional features present in the low temperature 
conductivity we 
plot in Fig.\ \ref{fig4} $\sigma/\sigma_b$ for U between 0.1 and 0.5\ V. The vertical 
lines 
correspond to the minima of the oscillations $U_m$ which are particularly visible at 
lower 
temperatures. To find the periodicity of these oscillations, we plot in Fig.\ \ref{fig5} 
$U_m$ vs. 
an integer index $n$. A clear zero crossing of the linear fit is obtained by choosing an 
index n=9 
for the lowest extracted value of $U_m$. From the linear fit we deduce a period of $(11.1 
\pm 
0.5)$\ mV. If we remember that this value corresponds to 8-10 junctions connected in 
series, a 
single junction would show a periodicity of $\delta U=(1.2\pm 0.1)$ meV.

The expected $c$-axis density of states of a superlattice where the superconducting gap 
is a one 
dimensional periodic step has been already calculated by van Gelder in 1969 
\cite{van_gelder}. 
With the HTS materials, the increasing interest on superlattices inspired the work of 
Hahn in 
extending the model to the three dimensional case \cite{hahn}. These models predict the 
opening 
of gaps which were particularly visible in the one dimensional case. In the 3-dimensional 
case 
they are smeared out, although still present. The main result from Ref.\ \cite{hahn} is 
that above 
the superconducting gap these additional gap structures should appear with a periodicity 
of: 
\begin{equation} 
\frac{\xi_c}{s}=\frac{1}{\pi^2} \frac{\delta U} {\Delta_c} 
\label{eq2} 
\end{equation} 
where $s$ is the periodicity of the superlattice, $\xi_c$ the $c$-axis coherence length, 
$\Delta_c$ 
the $c$-axis gap and $\delta U$ the periodicity of the sub-gap structures. By taking 
$s=10.5$\ 
nm, and $\Delta_c=5.0\pm 0.5$ \ meV we deduce from Eq.\ \ref{eq2} a $c$-axis coherence 
length 
of $\xi_c=0.27$ nm \cite{remark1}. This result is close to the value of $\xi_c=0.16\pm 
0.01$\ nm 
deduced from an analysis of fluctuation conductivity in Y123/Pr123 superlattices 
\cite{solovjov}. If we assume for Y123 an anisotropy of $\gamma \approx 5$ \cite{janossi} 
we 
would obtain a in-plane coherence length $\xi_{ab}=\gamma\xi_c\approx 1.4$\ nm. This 
value is 
close to the generally quoted $\xi_{ab}=1.5$\ nm for Y123 \cite{tinkham}. The larger 
structures 
indicated by the arrows in Fig.\ \ref{fig3}, correspond to a periodicity of $100\pm 1$ 
mV. It is 
interesting to notice that the ratio of this periodicity divided by the periodicity of 
$U_m$ is 
$9.0\pm 0.4$.This value corresponds to the ratio between $s$ and a Y123 unit cell! From 
Eq.\ 
\ref{eq2} and Fig.\ \ref{fig4} we deduce that $\xi_c$ is practically temperature 
independent 
below 20 K.

%

We show in this paper that by constructing superlattices it is possible to generate sub-
gap 
structures. These features can be directly used to determine in an independent way a $c$-
axis 
coherence length $\xi_c=0.24\pm 0.03$\ nm for a 2 u.c. thick Y123. 

The agreement of these results with previous estimations of Y123 coherence lengths shows 
that 
$\Delta_c=5.0\pm 0.5$ \ meV could be indeed a $c$-axis superconducting gap related to the 
CuO 
chains.

From the observed sub-gap structures, we conclude that $\xi_c(T)$ is practically 
temperature 
independence below 20\ K.

Due to thermal fluctuations, this analysis could not be extended up to higher 
temperatures.

The authors would like to thank K. Gray, J.F. Zasadzinski, R.A. Klemm, R. Schilling and 
J. 
Mannhart for valuable and stimulating discussions. This work was supported by the German 
BMBF through Contract 13N6916 and the European Union through the Training and Mobility 
of Researchers  (ERBFMBICT972217).

\newpage
%
%

\newpage
%
%

\newpage
%
%

\newpage
%
%

\newpage
%
%

\begin{references}
\bibitem{remark1} Since $\xi_c$ is here deduced from the ratio $\Delta_c/\delta U$, our 
results are 
independent from the number of junctions constituting the stack.
\end{references}
\end{document}